\documentclass[12pt]{iopart}
\usepackage{ragged2e}
\usepackage{graphicx, abstract, caption}
\usepackage{hyperref} % for hyperlinks
\expandafter\let\csname equation*\endcsname\relax
\expandafter\let\csname endequation*\endcsname\relax
\usepackage{amsmath}
\usepackage{mathptmx}
\usepackage{etoolbox}
\usepackage{parskip}
\usepackage{xcolor}
\usepackage{cite}
\usepackage{microtype}

\hypersetup{
    colorlinks=true,        % false: boxed links; true: colored links
    linkcolor=blue,         % color of internal links (sections, etc.)
    citecolor=blue,         % color of citation links
    urlcolor=blue           % color of external links
}

\usepackage{iopams}  
\begin{document}
\title{Exceptional point in a PT symmetric non-Hermitian terahertz plasmonic metasurface} 
\author{Anshul Bhardwaj\textsuperscript{1}, Maidul Islam \textsuperscript{2,*}, Chandan Kumar \textsuperscript{1}, Anuraj Panwar\textsuperscript{3},and Gagan Kumar\textsuperscript{1}}

\address{\textsuperscript{1}Department of Physics, Indian Institute of Technology Guwahati, Guwahati, Assam 781039, India}
\address{\textsuperscript{2}
Department of Physics, The Assam Royal Global University, Guwahati, Assam 781035, India}
\address{\textsuperscript{3}Department of Physics and Materials Science and Engineering, Jaypee Institute of Information Technology, Noida-62, Uttar Pradesh 201309, India}

\ead{\textsuperscript{*}mislam@rgu.ac} 

\begin{abstract}
In this paper, we experimentally demonstrate a non-Hermitian open PT-symmetric terahertz metasurface comprising complementary plasmonic structures capable of exhibiting an exceptional point (EP). The metasurface consists of two resonators of different sizes, representing effective gain and loss elements, placed orthogonally in close proximity to realize a non-Hermitian configuration leading to a PT symmetry state. A diagonal displacement of one resonator within this strongly coupled near-field configuration leads to the emergence of an exceptional point, where the system undergoes a sudden phase transition from a PT-symmetric to a PT-asymmetric state. Terahertz time-domain spectroscopy (THz-TDS) is performed on the fabricated samples to experimentally validate the phase transition observed in numerical simulations. We employ coupled mode theory (CMT) to analyse and distinguish between the PT-symmetric, exceptional point, and PT-asymmetric states. This theoretical framework enables the calculation of eigenvalues, phase spectra, and eigenmodes associated with the metamaterial design, thereby corroborating the simulation results. Furthermore, we construct Poincaré sphere to visualize the orientation of the polarization states of the eigenmodes, which further indicates the presence of the exceptional point. This comprehensive study of exceptional point in a plasmonic system holds potential for the development of practical, highly sensitive terahertz devices, addressing limitations of conventional PT-symmetric systems that rely on traditional gain and loss media.

\end{abstract}
(*Corresponding author)
\section{\label{sec:level1}Introduction}
The concept of parity-time (PT) symmetry introduced by Bender et al. has been one of the remarkable findings in quantum mechanics and has facilitated the open non-Hermitian physical system to possess real eigenvalues \cite{bender1998real}. Since the discovery of PT symmetry, it has become an active research area, and a tremendous progress has been made both in theory and experiments in various systems including atomic-quantum systems \cite{zhang2016observation, gao2015observation}, electronics \cite{schindler2011experimental}, acoustics \cite{fleury2015invisible, zhu2014pt}, opto-mechanics\cite{jing2015optomechanically}, etc. Most recently, the concept of PT symmetry has intrigued photonics researchers, especially working in metamaterials\cite{ozdemir2019parity}. The potential energy of the metasurface remains invariant under the application of the combined parity $\hat{\text{P}}$ and time-reversal $\hat{\text{T}}$ operators. Despite the different losses viz. diffraction, scattering, and propagation loss, the open metamaterials systems can exhibit PT symmetry with the design of lossy and gain media. For instance, Zhu et al. investigated asymmetric diffractions between a pair of oblique incident light in a one-dimensional parity-time symmetric diffraction grating with balanced gain and loss media \cite{zhu2016asymmetric}. Chang et al. experimentally reported parity–time symmetry using a chip at 1,550 nm wavelength in two directly coupled microtoroid resonators maintaining balanced gain and loss media \cite{chang2014parity}. The key concept of parity-time symmetry is integrated with the observation of the exceptional point (EP) of the non-Hermitian metasurface design, where the system suddenly becomes PT asymmetric from symmetric one. The degenerate eigenvalues and coalescence of eigenstates at EP ensures intriguing phenomena viz chiral response, vortex generation, loss-induced transparency \cite{zhang2021chiral, leung2020exceptional, guo2009observation}. The most of the work in these areas has been reported in the optical or microwave regions.\\
The terahertz region (0.1 – 10 THz) of the electromagnetic spectrum has been given attention in exploring parity-time symmetry and exceptional point by maintaining balanced gain and loss media. In this regard, Lawrence et al. investigated exceptional point in parity-time symmetric metasurface comprised C-shaped resonators of different metal with different conductivities acting like gain and loss media \cite{lawrence2014manifestation}. Sakhdari et al. reported wave manipulation and highly sensitive sensor using PT symmetric metasurface \cite{sakhdari2017pt}. Wang et al. explored parity-time symmetry and exceptional point in a traditional metasurface comprising two orthogonally arranged split ring resonators \cite{wang2021active}. Fu et al. reported third order exceptional point in THz metasurface by tuning near-field coupling between three gold split-ring resonators \cite{fu2025achieving}. Based on near-field coupling mechanism in cross-polarization channels, the study of exceptional point has been also examined for bio-sensors, fano-resonance, superconducting phase transition and topological properties etc \cite{xu2022experiment, liang2024engineering, wang2017superconductive, zhang2025terahertz, li2020exceptional}.\\
The study of parity-time symmetry and exceptional point has been carried out using traditional metamaterial constituents of different conductivities representing the gain and loss media on a dielectric substrate. At terahertz frequencies, this poses a significant challenge in experiments as metals behave like a perfect conductor.   In our work, we have tried to address it by proposing a novel non-Hermitian parity-time symmetric terahertz metasurface. The unit cell of the metasurface comprises orthogonally placed complementary resonators of unequal sizes. The smartly designed resonators, placed in a near-field configuration, act as gain and loss media, and exhibit parity-time symmetry. The physical displacement of the resonator and novel approach reported in our work enable the observation of the experimental point overcoming the limitations of traditional loss-gain media. The novel findings on highly sensitive exceptional points open new avenues for building practical devices at terahertz frequencies.\\
The article has been organized as follow: Section 2 presents the design and experimental details of the proposed non-Hermitian fabricated THz plasmonic metasurface with the measurement process. In Section 3, we examine the key results associated with the fabricated metasurfaces. The simulation and experimental results also have been reported in this section. Specifically, section 3.1 outlines the semi-analytical coupled mode theory used to analyse PT symmetric phase transition and its exploitation via an exceptional point. This section also includes theoretical transmission results as a function of the diagonal displacement between the centres of the two resonators. In Section 3.2, we discussed the eigenvalues, phases, and eigenstates to further investigate PT symmetry breaking and the presence of the EP in the metasurface. Section 4 explores Poincaré sphere and circular polarization to characterize the non-Hermitian nature of the meta-design. Finally, Section 5 concludes the article by summarizing the main findings.

\section{Design of PT symmetric terahertz metasurface and experimental details}
\begin{figure}[!ht]
	\centering
	\includegraphics[width=1\linewidth]{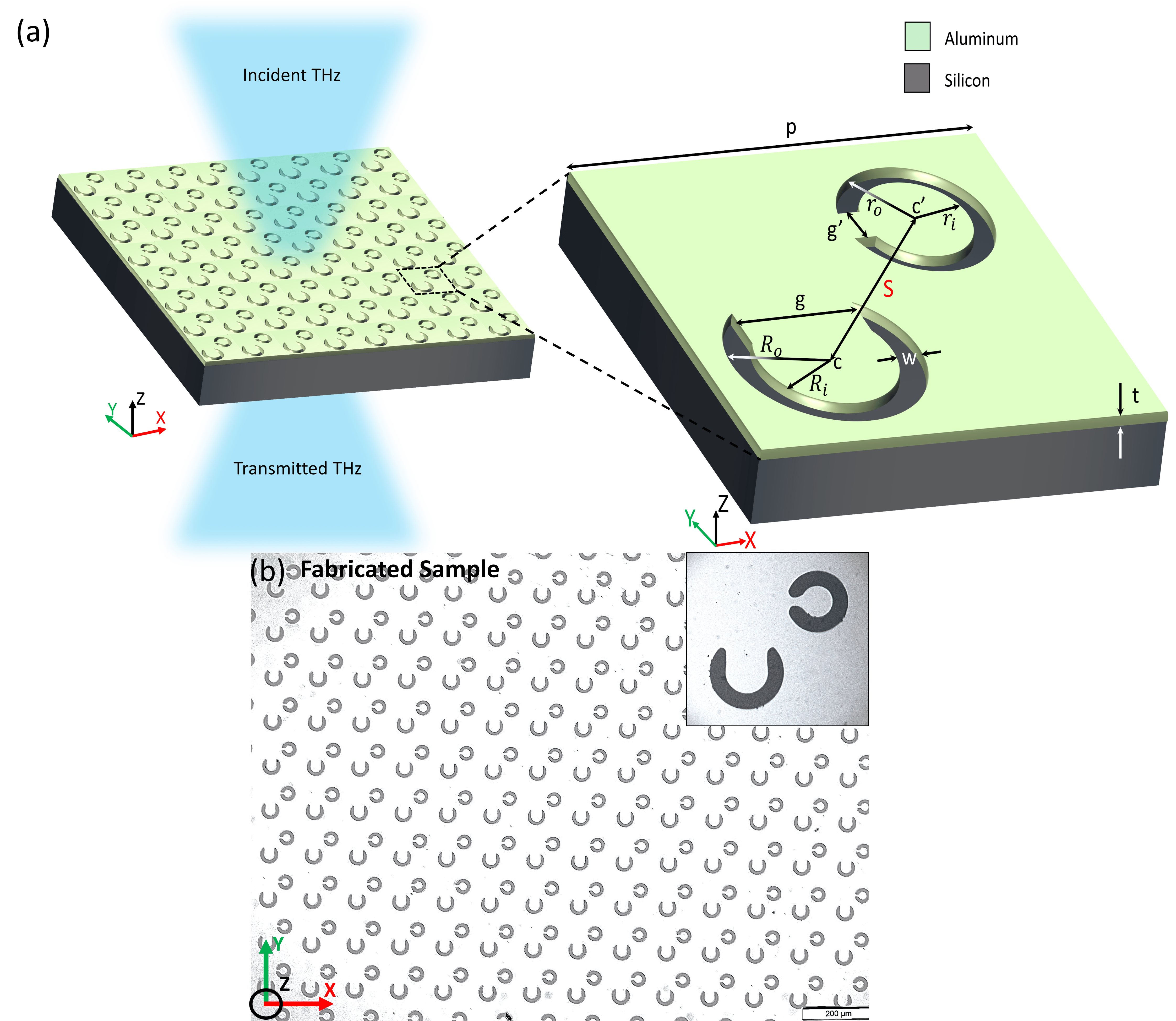}
	\caption{\footnotesize(a) Schematic of the proposed 2D THz plasmonic metasurface consisting of periodically arranged different size circular etched cavities on silicon substrate in the x- and y-directions, the unit cell of metasurface structure. The resonators are separated by the inter-cavity separation displacement (S). The parameters c, g, R${_\text{o}}$ and R${_\text{i}}$ represent the center, split gap, outer \& inner radius of the bigger cavity, while c', g', r${_\text{o}}$ and r${_\text{i}}$ indicate center, split gap, outer \& inner radius of smaller cavity, respectively. The thickness of the Al-thin film, width of the cavities and periodicity of the unit-cell are denoted by t, w and p, respectively. (b) Optical microscopic view of the fabricated 2D THz plasmonic metasurface consisting of periodically arranged circular shaped cavities along x- and y- directions with unit cell of fabricated sample.}
	\label{f:1}       
\end{figure}
We propose a fabricated non-Hermitian terahertz (THz) plasmonic metamaterial design to investigate parity-time (PT) symmetry breaking and the emergence of an exceptional point (EP) associated with eigenstate phase transition. The design comprises a two-dimensional non-Hermitian array of periodically arranged unit cells, each consisting of orthogonally oriented, circular-shaped complementary split-ring resonators of two different sizes, etched with distinct split gaps on a silicon substrate. The schematic of the proposed 2D non-Hermitian THz plasmonic meta-design can be seen in Fig. \ref{f:1}(a) with a unit cell with periodicity p=125 µm. The metal layer is composed of aluminium (Al), with a conductivity of 3.56$\times$10$^{7}$ S/m and a thickness t=0.200 µm. The square-shaped split gap of the bigger cavity (g= 28 µm) is oriented perpendicular to that of the smaller cavity (split gap g'= 6.5 µm). The outer (R${_\text{o}}$) and inner (R${_\text{i}}$) radii of the bigger cavity are 24 µm and 16 µm; while the smaller cavity has outer (r${_\text{o}}$) and inner (r${_\text{i}}$) radii of 20 µm and 12 µm, respectively. The variation in cavity sizes introduce effective gain and loss elements within the meta-atoms. The width (w) of both cavities is uniformly set to 8 µm. The position of the bigger cavity, along with unit cell parameters, is kept constant during the investigation of the PT-symmetric phase transition in the proposed THz plasmonic meta-device. The centre-to-centre (c–c') inter-cavity separation distance (S) between the two cavities within the unit cell is systematically varied to examine this transition. Five different values of S are considered: S${_\text{1}}$= 64 µm, S${_\text{2}}$= 66 µm, S${_\text{3}}$= 68 µm, S${_\text{4}}$= 70 µm and S${_\text{5}}$= 72 µm, using a step size of 2 µm. This variation is intended to explore the emergence of the exceptional point marking the phase transition from PT-symmetric to PT-broken states.\\
Considering the above-mentioned parameters, the samples were fabricated using a photolithography technique on a silicon substrate in a clean room environment. The substrate was first cleaned sequentially with acetone, isopropyl alcohol, and finally distilled water. A 0.2 µm thick aluminum thin film was deposited onto the substrate using thermal evaporation. A layer of positive photoresist (S1813) was spin-coated on the Al-film. The metasurface design was patterned onto the photoresist using a mask writer. After exposure, the sample was treated with a developer solution to remove the exposed areas, revealing the underlying Al-film. The exposed aluminum was selectively etched using an Al-etchant, forming the desired metasurface structure. The remaining photoresist was removed by rinsing the sample with acetone. Optical microscope image of the fabricated sample is shown in Fig. \ref{f:1}(b) with the etched cavities on Al-film attached on substrate with unit-cell in square-box.

\begin{figure}[!ht]
	\centering
	\includegraphics[width=1\linewidth]{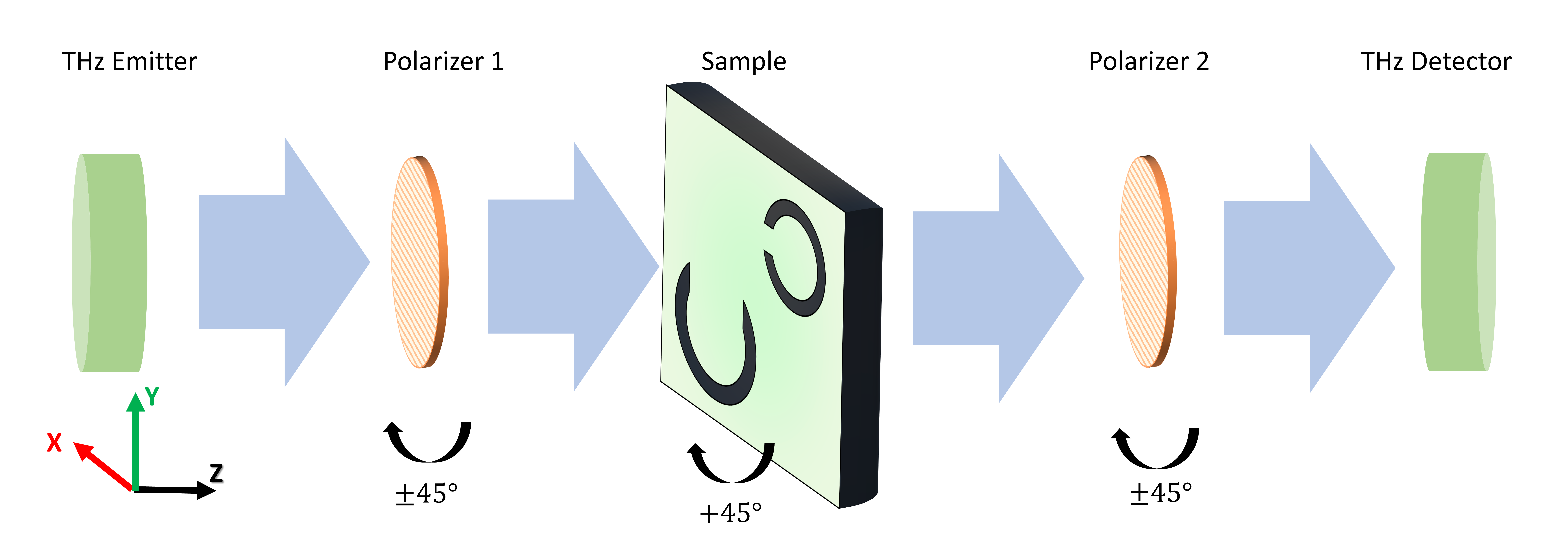}
	\caption{\footnotesize Terahertz characterization of fabricated THz plasmonic metasurface. THz wave propagation is along z-direction. The co- and cross-polarization transmission components can be examined using two polarizers}
	\label{f:2}       
\end{figure}
The transmission properties of the proposed fabricated non-Hermitian THz plasmonic metasurface can be investigated using THz Time-domain Spectroscopy (THz-TDS). A schematic of the experimental setup is illustrated in Fig. \ref{f:2}. In this configuration, the electric field of the incident THz wave is aligned along the x-direction, while the wave propagates along the z-direction. With the help of two polarizers $\text{T}_\text{xx}$, $\text{T}_\text{xx}$, $\text{T}_\text{xy}$ and $\text{T}_\text{yx}$ transmission components results can be extracted from the THz-time domain spectroscopy setup. To measure the co- and cross-polarization transmission characteristics, initial calibration is performed using a silicon substrate with two polarizers. The co-polarization and cross-polarization responses are obtained by aligning the polarizers in the same and perpendicular orientations (i.e. at ±45°), respectively. Next, the fabricated THz plasmonic metasurface is fixed at a +45° angle in the clockwise direction, while both Polarizer 1 and Polarizer 2 are rotated in the same direction to measure the co-polarization transmission components ($\text{T}_\text{xx}$ and $\text{T}_\text{yy}$). The cross-polarization transmission amplitudes ($\text{T}_\text{xy}$ and $\text{T}_\text{yx}$) are measured by orienting the polarizers in opposite directions. The measured time-domain signals are transformed into frequency-domain data using Fast Fourier Transform (FFT). The final transmission spectra as a function of frequency (in THz) are obtained by taking the ratio of the measured transmission through the fabricated sample to that through the silicon substrate, which serves as the reference. 

\section{PT symmetry in terahertz plasmonic metasurface: Results and Discussion}
Numerical simulations were performed using the commercially available software CST Microwave Studio™, employing the finite-difference time-domain (FDTD) method to investigate the complex transmission properties of a plasmonic metasurface structure. The design consists of a two-dimensional periodic array of orthogonally oriented circular cavities of different sizes. A hexahedral mesh was utilized in the frequency domain solver, with unit cell boundary boundary conditions along the x- and y-directions, respectively. The incident linearly polarized x- and y-polarized THz waves interact with the cavities. \begin{figure}[!ht]
	\centering
	\includegraphics[width=0.7\linewidth]{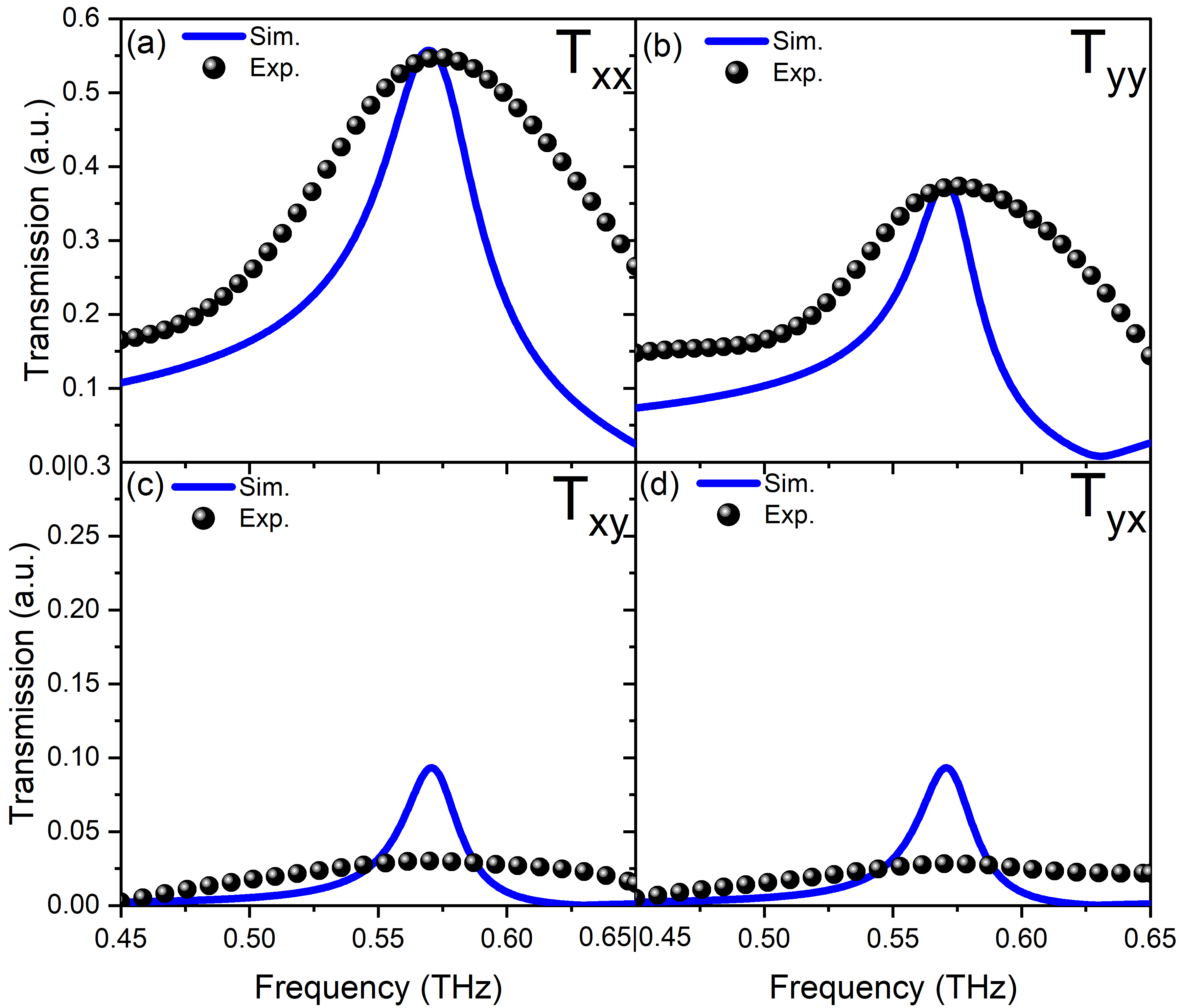}
	\caption{\footnotesize Simulation and experimental transmission properties in polarization space for S${_\text{3}}$= 68 µm. (a) and (b) represent the co-polarization transmission $\text{T}_\text{xx}$ and $\text{T}_\text{yy}$ with same resonance frequency, respectively. On the other hand, (c) and (d) depict the cross-polarization transmission $\text{T}_\text{xy}$ and $\text{T}_\text{yx}$, respectively. The blue colour and black bubbles are indicating the simulation and experimental results, respectively.  }
	\label{f:3}       
\end{figure}These cavities, structurally resembling complementary split-ring resonators (CSRRs), are near-field coupled in cross-polarization, characterized by the coupling coefficient $\Omega$. The complex transmission response of the proposed metasurface is probed at the bottom surface along the z-direction. To explore the PT-symmetric characteristics of the structure, the two cavities were initially placed at a centre-to-centre separation of S${_\text{1}}$ = 64 µm. Subsequently, the smaller cavity was laterally displaced along the axis of separation to new positions corresponding to S${_\text{2,3,4,5}}$ = 66, 68, 70, and 72 µm, to examine the evolution of PT symmetry and coupling dynamics.
To verify the resonance frequency for the fabricated sample, the comparison of the simulation and experimental transmission measurements for S${_\text{3}}$= 68 µm can be found in Fig. \ref{f:3}. Each experimental transmission polarization component not only shares the common resonance frequency with the simulation but also shows identical cross-polarization features, as seen in Fig. \ref{f:3}(c) and (d).\\
To investigate the evolution of PT-symmetric phase transition in the plasmonic THz non-Hermitian metasurface design, numerical simulations were carried out by orthogonally displacing the smaller cavity. The complex transmission properties are examined for various values of the centre-to-centre separation S in polarization space. The numerically and experimentally obtained transmission amplitudes $\text{T}_\text{xx}$, $\text{T}_\text{yy}$, $\text{T}_\text{xy}$ and $\text{T}_\text{yx}$ are presented in Fig. \ref{f:4}(a)–(d), respectively, for different S values.\\
It is observed that the co-polarized components $\text{T}_\text{xx}$ and $\text{T}_\text{yy}$ exhibit plasmonic resonances around 0.57 THz and different transmission amplitude is attributed to the unequal dissipation losses in the two cavities. The identical cross-polarized components $\text{T}_\text{xy}$ and $\text{T}_\text{yx}$, show a decreasing transmission amplitude with increasing S values, particularly around the resonance frequency. On the other hand, a similar trend is followed by experimentally obtained co- and cross transmission spectra, as shown in Fig. 4(e-h). In Fig. 4(e) and (h), the experimentally observed co-polarization transmission components $\text{T}_\text{xx}$, $\text{T}_\text{yy}$ and cross polarization components $\text{T}_\text{xy}$ and $\text{T}_\text{yx}$ are exhibitting resonance frequency around 0.57 THz. Further the experimentally measured amplitudes of different components $\text{T}_\text{xx}$, $\text{T}_\text{yy}$, $\text{T}_\text{xy}$ and $\text{T}_\text{yx}$ are found to be decreased with increases the increase in inter-cavity separation displacement S.
%%%%%%%%%%%%%%%%%%%%%%%%%%%%%%%%%
\begin{figure*}[!ht]
    \centering
    \begin{minipage}{0.5\linewidth}
        \centering
        \includegraphics[width= \linewidth]{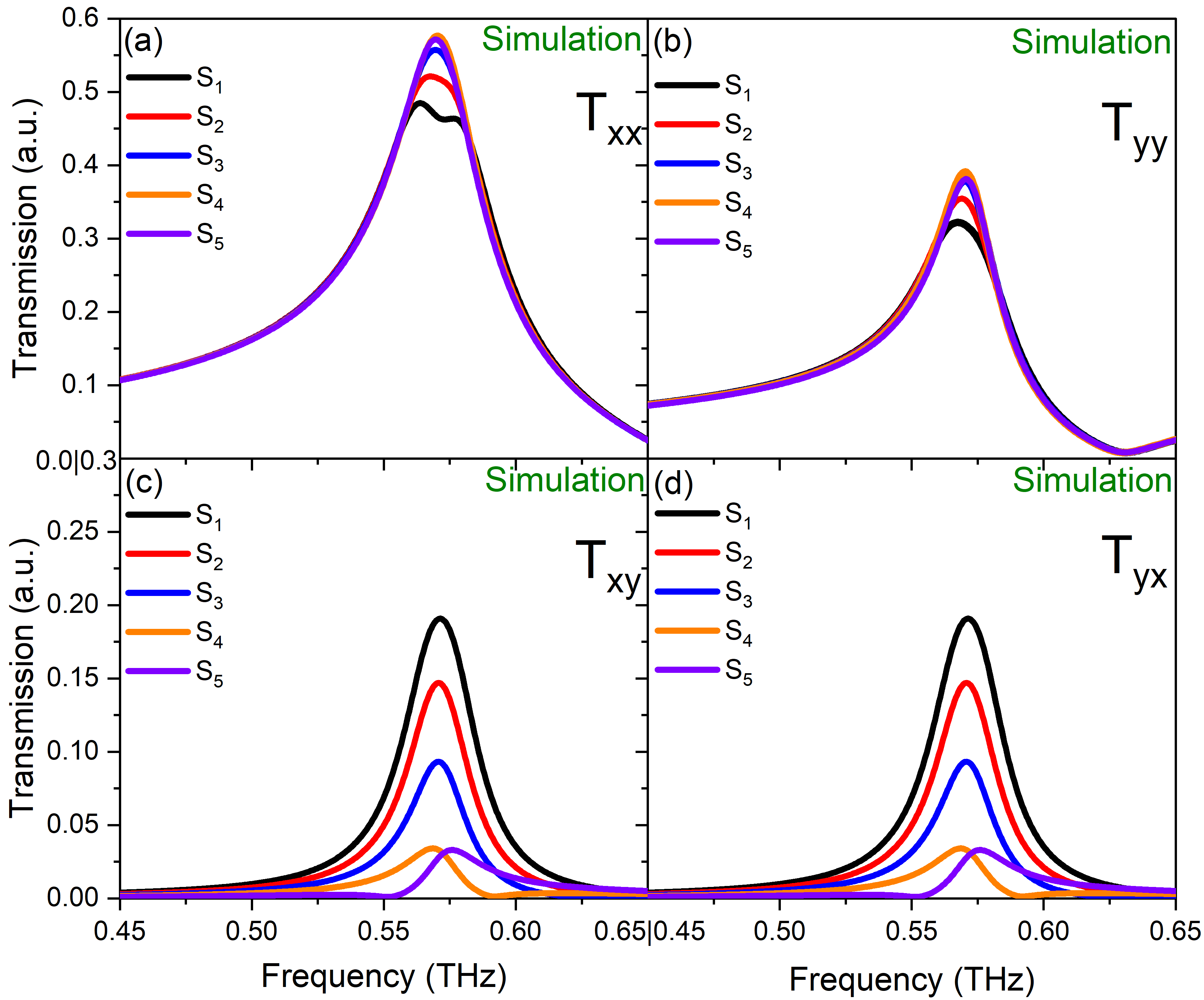}
    \end{minipage}\hfill
    \begin{minipage}{0.5\linewidth}
        \vspace{2mm}
        \centering
        \includegraphics[width=\linewidth]{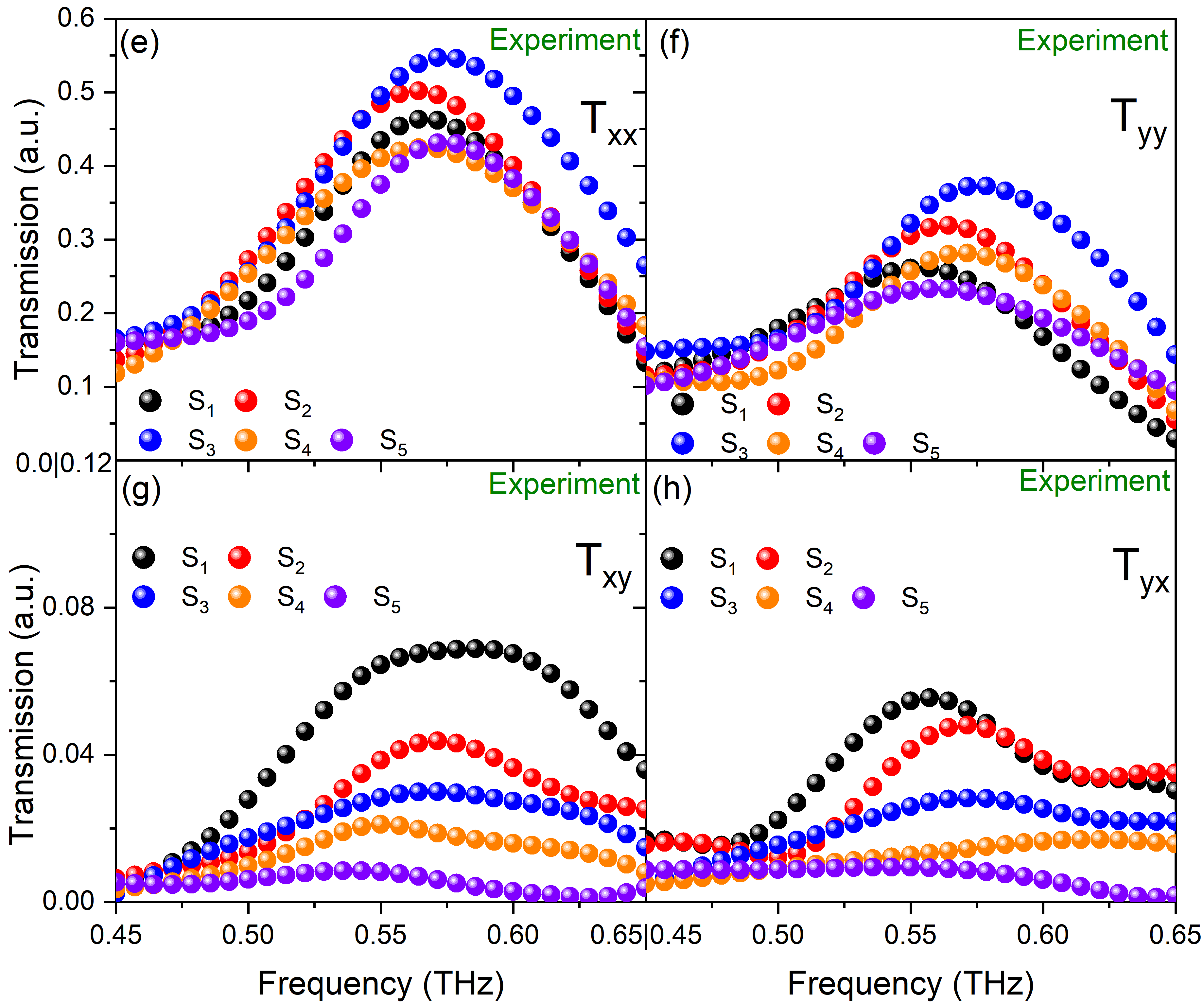}
    \end{minipage}
    \caption{\footnotesize Simulation and experimental results for co- and cross- polarization T-matrix components in the proposed THz plasmonic metasurface design for separation distance S${_\text{1,2,3,4,5}}$= 64, 66, 68, 70 and 72 µm are indicated through different color lines, respectively. Figures (a), (b), (c) and (d) represent numerically obtained transmission amplitudes $\text{T}_\text{xx}$ ,$\text{T}_\text{yy}$, $\text{T}_\text{xy}$ and $\text{T}_\text{yx}$ of metasurface geometry for different S values for x and y-polarization of incident THz wave, respectively. Similarly, (e-d) depict experimentally obtained the same T-matrix components of fabricated THz metasurfaces, respectively.}
    \label{f:4}
\end{figure*}
\subsection{Semi-analytical approach based on coupled mode theory:}
Semi-analytical coupled mode theory (CMT) has been employed to understand the physical insights behind the PT symmetry nature of the proposed THz plasmonic metasurface geometry along with  the coupling mechanism of complementary resonators under the effect of the incident THz wave. The diagonally placed cavities of the non-Hermitian metasurface depend on the polarization of the incident THz wave. The proposed design exhibit plasmonic resonances, supported by the two cavities under the interaction with the incident THz beam can be explained through the two-time dependent complex mode equations shown below based on coupled mode theory\cite{fan2003temporal, zhu2013temporal, li2022bifunctional, li2020exceptional}, 
\begin{align}
\frac{d\Tilde{a}_{x,y}}{dt} &=(i{\omega}_{x,y}-\xi_{x,y}) \Tilde{a}_{x,y}+i\Omega \Tilde{a}_{y,x}+i \sqrt{\gamma_x,y}E_{x,y}^{in},
\end{align} 
On the other hand, corresponding the output of the electric fields in polarization space can be written as,
\begin{align}
\begin{bmatrix}
E_x^{out} \\
E_y^{out} 
\end{bmatrix}
&\equiv \begin{bmatrix}
i\sqrt{\gamma_x} & 0\\
0 & i\sqrt{\gamma_y} 
\end{bmatrix} \begin{bmatrix}
\Tilde{a}_x\\
\Tilde{a}_y
\end{bmatrix}= \widetilde{\text{T}}\begin{bmatrix}
E_x^{in} \\
E_y^{in} 
\end{bmatrix},
\end{align} 
Where, E$^{\text{out,in}}_{x,y}$ represent the transmitted and incident electric field components for x- and y-polarizations, respectively. The parameters $\omega_{x,y}$, $\gamma_{x,y}$, $\Gamma_{x,y}$ and $\Tilde{a}_{x,y}$  represent the resonance frequencies, radiative losses, dissipation losses and complex resonance modes for the x- and y-polarized THz waves, respectively. The total loss, comprising both radiative and dissipative components, is expressed as ($\gamma_{x,y}$+$\Gamma_{x,y}$)= $\xi_{x,y}$ for the respective polarization.  In order to examine the numerical properties, $\widetilde{T}$ represents the complex transmission jones matrix of the system. Based on above written two set equations, PT symmetric non-Hermitian Hamiltonian can be expressed as with $\gamma_x\delta_y-\gamma_y\delta_y=0$ condition (detuning frequency $\delta{_x}{_,}{_y}$= ($\omega - \omega{_x}{_,}{_y}$)),
\begin{align}
\text{H}_\text{o} &= \begin{bmatrix}
-\gamma_y \delta_x-i \eta\sqrt{\gamma_x} \sqrt{\gamma_y} && \Omega \sqrt{\gamma_x} \sqrt{\gamma_y} \\
\Omega \sqrt{\gamma_x} \sqrt{\gamma_y} &&  -\gamma_x \delta_y+i\eta\sqrt{\gamma_x} \sqrt{\gamma_y}
\end{bmatrix}. 
\end{align}
Now, the new reference point is the effective loss of the system, which can be written in terms of,
\begin{align}
\eta &= {\frac{(\gamma_x\Gamma_y - \gamma_y\Gamma_x)}{2 \sqrt{\gamma_x}\sqrt{\gamma_y}}}.
\end{align}
To preserve PT symmetry, the Hamiltonian $H_o$ of the non-Hermitian system must commute with the parity-time operator ($\hat{P}\hat{T}$), i.e., [$\hat{P}\hat{T},H_o$]= 0. Using Eq. (1-2), the complex transmission jones matrix $\widetilde{T}$ can be expressed as,
\begin{align}
|{\Tilde{\text{T}}}| &= \left|\frac{i \gamma_x \gamma_y}{\det(i \chi \text{I} +\text{H}_\text{o})} (\chi \text{I} - \text{adj}(\text{H}_\text{o}) )\right|.
\end{align}
Where, $\chi=({\gamma_x\mathrm{\Gamma}}_y+\gamma_y\mathrm{\Gamma}_x)/2\ +\gamma_x\gamma_y$ and I is 2$\times$2 unit matrix.
Furthermore, PT symmetry and exceptional point (EP) of the non-Hermitian system can be analysed using the following expressions for the eigenvalues and eigenstates of the T-matrix, respectively.
\begin{align}
\lambda_\text{T}= (\lambda_{\text{H}_\text{o}} + i \chi)\bigg(\frac{\alpha-i \beta}{\alpha^2+\beta^2}\bigg),
\end{align}
where \begin{align}
\lambda_{\text{H}_\text{o}} &= \bigg[ \frac{(-\gamma_x \delta_y - \gamma_y \delta_x)}{2} \pm \frac{\sqrt{\left[(\gamma_x \delta_y - \gamma_y \delta_x) - 2i \eta \sqrt{\gamma_x} \sqrt{\gamma_y} \right]^2 + 4 \gamma_x \gamma_y \Omega^2}}{2} \bigg]
\end{align}
Where $\alpha= (\xi_x \delta_y + \xi_y \delta_x)$ and $\beta= (\delta_x \delta_y - \xi_x \xi_y -\Omega^2)$. At the eigen-frequency, equations (6-7) describe PT symmetric phases and the exceptional point of the system through the following three distinct cases: Case I: When the coupling is stronger than the losses of the cavities $|\eta|<|\Omega|$, the non-Hermitian metasurface operates in the PT-symmetric phase. The eigen-polarization states are given by ($\hat{x}$, $\pm$ $\hat{y}$ $exp(\pm(i\theta))$), where $\theta= \sin^{-1}(\eta/\Omega)$. In this regime, the eigenstates are oriented at ±45°. Case II: For the opposite case, when the coupling is weaker than the losses $|\eta|>|\Omega|$, the metasurface transition into the PT-broken phase. The corresponding eigen-polarization states are ($\hat{x}$, $\hat{y}$ $sgn(\eta/\Omega) i exp(\pm\theta)$), where $\theta= cosh^{-1}(|\eta/\Omega|)$ and sgn denotes the sign function. Here, the eigenstates align along 0° and 90° axes. Case III: At one point, the coupling and effective loss of the system are equal $|\eta|=|\Omega|$, the system reaches the exceptional point (EP), marking a sharp transition from the PT-symmetric to the PT-broken phase. At this point, the eigenvalues of the transmission matrix $\widetilde{T}$ and the system PT-symmetric non-Hamiltonian and H$\text{o}$ coalesce, and the corresponding eigen-polarization states become degenerate circular states, expressed as ($\hat{x}$, $\hat{y}$ $i sgn(\eta/\Omega)$). This condition characterizes the exceptional point where PT symmetry is effectively exploited.\\
In order to understand the nature of PT symmetric design, Coupled Mode Theory (CMT) has been applied to support the numerically obtained T-matrix components shown in Figs. 3 and 4. Eq. (5) is used to calculate the various components of the T-matrix. The theoretically derived co- and cross-polarization transmission components are plotted in Figs. 5(a)–(d). These theoretical results show good agreement with the numerical transmission spectra presented in Figs. 3 and 4. The variation observed in the cross-polarization transmission components $\text{T}_\text{xy}$ and $\text{T}_\text{yx}$ arises due to the coupling coefficient $\Omega$ between the cavities of the non-Hermitian metasurface design. \begin{figure*}[!ht]
    \centering
    \begin{minipage}{0.5\linewidth}
        \centering
        \includegraphics[width= \linewidth]{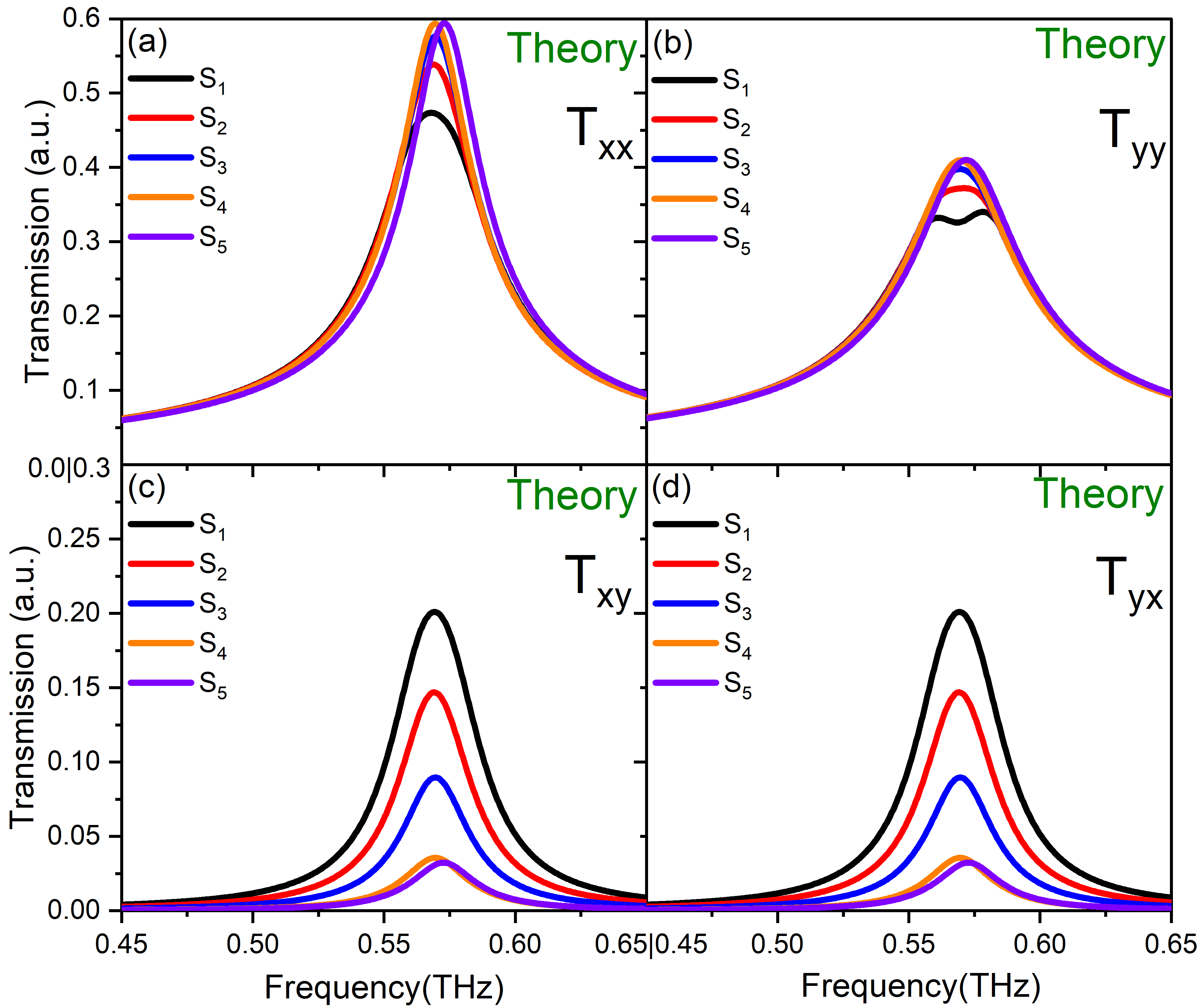}
    \end{minipage}\hfill
    \begin{minipage}{0.48\linewidth}
        \vspace{0.5mm}
        \centering
        \includegraphics[width=\linewidth]{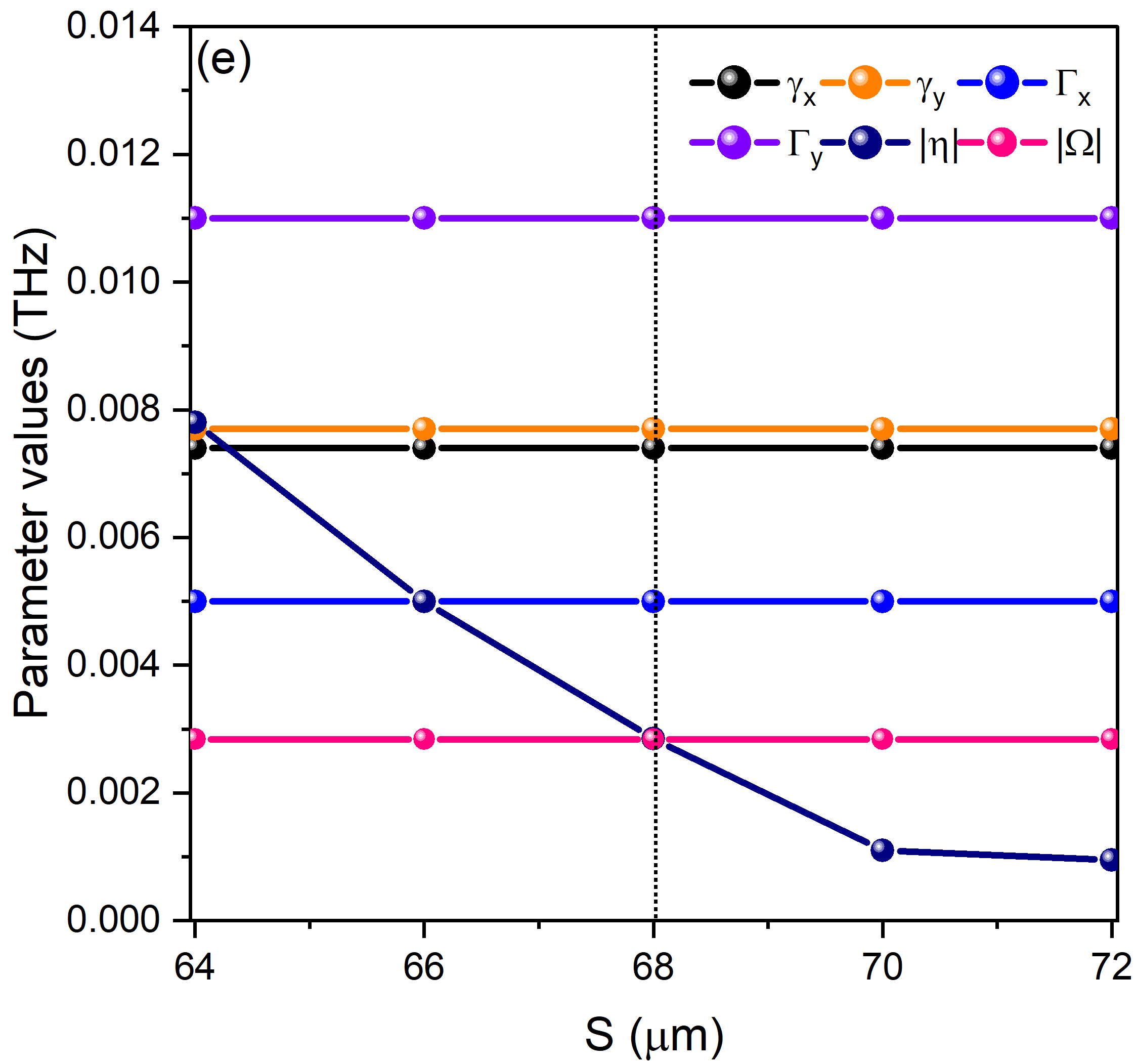}
    \end{minipage}
    \caption{\footnotesize Theoretically obtained co- and cross-polarization transmission spectra of the THz plasmonic metasurface for varying values of S. Figures (a)–(d) show the transmission components components $\text{T}_\text{xx}$, $\text{T}_\text{yy}$, $\text{T}_\text{xy}$ and $\text{T}_\text{yx}$, respectively. Figure (e) presents the corresponding fitted parameters used to obtain the transmission results of the meta-design. The radiative losses and dissipation losses are denoted by $\gamma_{x,y}$ and $\Gamma_{x,y}$ for the x- and y-polarized THz waves, respectively. The coupling coefficient $\Omega$ and the effective loss $\eta$ intersect at $\text{S}_\text{3}$= 68 µm, marked by a dashed line, which indicates exceptional point.}
    \label{f:5}
\end{figure*}
The resemblance of theoretical spectra with the numerical spectra is due to the specific values of different parameters involved in coupled mode theory, which are shown in Fig. \ref{f:5}(e). It is evident that the coupling coefficient $\Omega$ decreases with increasing centre-to-centre inter-cavity separation distance S between the two complementary split ring resonators. This decline in $\Omega$ is attributed to reduce near-field coupling in the cross-polarization channels, while the remaining parameters remain constant. 
\subsection{Phase transition through exceptional point}
\begin{figure}[!ht]
	\centering
	\includegraphics[width=1\linewidth]{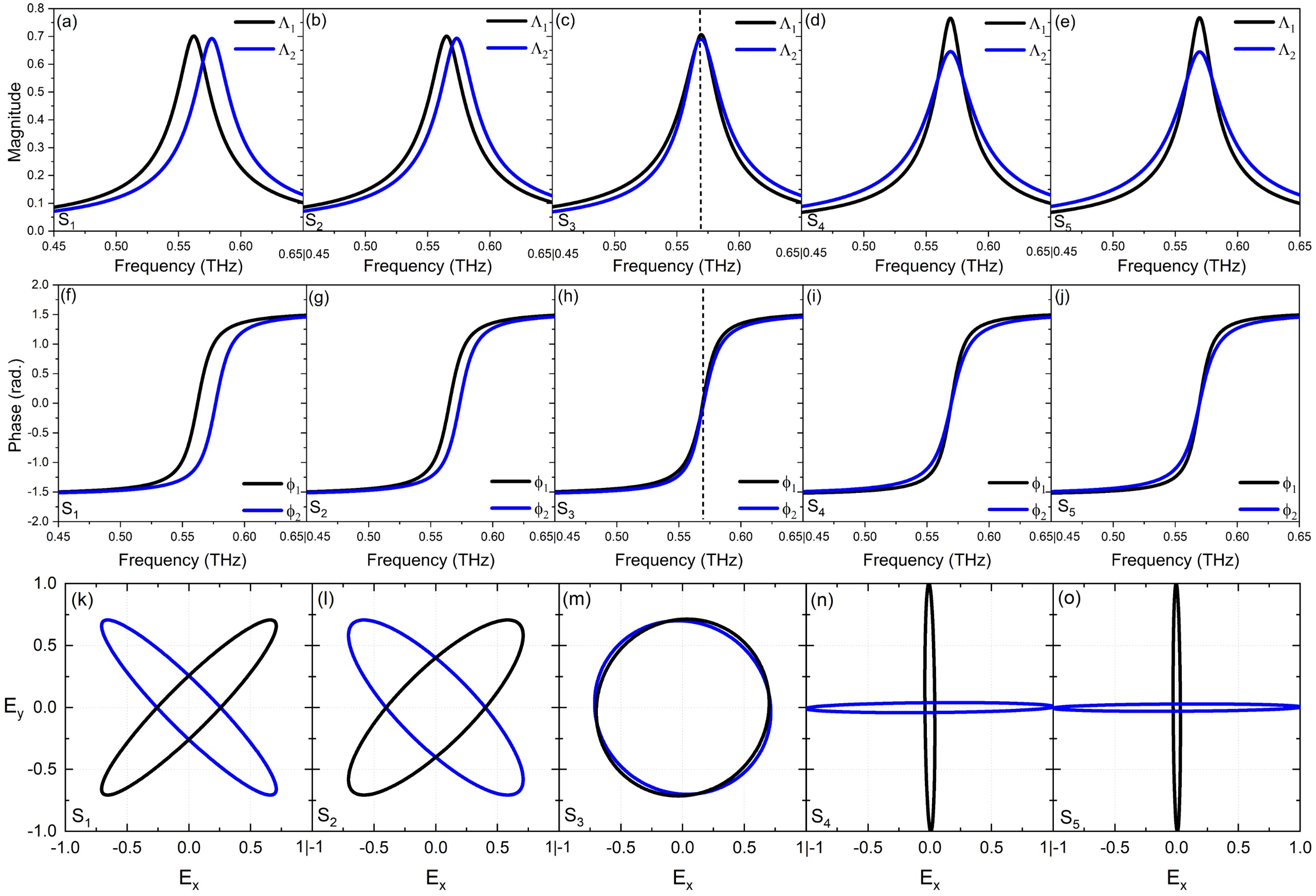}
	\caption{\footnotesize Here, $\Lambda{_1,_2}$ denote the magnitudes of two eigenvalues supported by the plasmonic metasurface. Figures 6(a,b), (f,g), and (k,l) represent the unbroken PT-symmetric phase of the meta-design. Figure 6(c), (h), and (m) correspond to the exceptional point (EP) at 0.567 THz, where the plots become nearly identical. In contrast, Figs. 6(d,e), (i,j), and (n,o) indicate the broken PT-symmetric phase of the meta-structure.}
	\label{f:6}       
\end{figure}Further, we investigate parity-time symmetry breaking through the exceptional point in the proposed THz plasmonic metasurface design. In order to do this, Eigenvalues, eigen-phase spectra, and eigenstates are calculated using the fitted parameters from coupled mode theory, as shown in Fig. \ref{f:5}(e), for different separation values S${_\text{1,2,3,4,5}}$ = 64, 66, 68, 70, and 72 µm, respectively. The corresponding eigenvalues and eigen-phase spectra are shown in Fig. \ref{f:6}(a)–(e) and Fig. \ref{f:6}(f)–(j), respectively, while the eigenpolarization states are presented in Fig. \ref{f:6}(k)–(o).
As seen in Fig. \ref{f:6}(a) and (b), the metasurface exhibits real and distinct eigenvalues at S${_\text{1}}$ = 64 µm and S${_\text{2}}$ = 66 µm. Similar trends are observed in the phase spectra at these separations, shown in Fig. \ref{f:6}(f) and (g), indicating the system is in the PT-symmetric phase or unbroken PT symmetry phase. This is further validated by the eigenpolarization states in Fig. \ref{f:6}(k) and (l) for S${_\text{1,2}}$, where the polarization ellipses are oriented at ±45°, confirming the PT-symmetric nature due to strong near-field coupling. It is observed that the minor axes of both co-rotating ellipses increase as the coupling strength $\Omega$ decreases.\\
Next, the smaller cavity is further shifted and centre to centre separation displacement value set at S${_\text{3}}$ = 68 µm. The corresponding eigenvalues, phase spectra, and eigenstates are shown in Fig. \ref{f:6}(c), (h), and (m), respectively. As observed, the eigenvalue and eigenphase spectra nearly identical and exhibit a common eigenfrequency around 0.567 THz, indicated by the vertical dashed line. At this point, the coupling between the cavities due to THz wave interaction and the effective loss of the system become nearly balanced. In other words, the magnitude of the coupling and the effective loss satisfy the condition $|\Omega|$$\approx$$|\eta|$. Furthermore, the co-rotating polarization eigenstates confirm the presence of an exceptional point at the eigenfrequency of around 0.567 THz in the non-Hermitian THz metasurface. At this point, the minor axes of both ellipses are nearly degenerate, forming two co-rotating circular polarization ellipses. This corresponds to a vortex state of the system, characterized by nearly degenerate eigenstates. After this point, the metasurface transition from the PT-symmetric to the PT-asymmetric phase, marked by a sudden change in the orientation angles of the eigenpolarization states. Subsequently, the inter-cavity separation distances are further increased to S${_\text{4}}$ = 70 µm and S${_\text{5}}$ = 72 µm, and the corresponding eigenvalues, phase spectra, and eigenstates are shown in Fig. \ref{f:6}(d–e), (i–j), and (n–o), respectively. The eigenvalue spectra are nearly identical at an almost single frequency but exhibit different linewidths, while the eigen-phase spectra intersect with each other at same frequency. The eigenstates at these separations are oriented along the x- and y-axes, forming a 90° angle between them. As we can see, the sudden orientation angles phase change is observed after exceptional point. These features indicate the onset of PT symmetry breaking in THz plasmonic metasurface. At S${_\text{4,5}}$ = 70 and 72 µm, the metasurface enters the PT-asymmetric phase, characterized by orthogonally aligned cavity resonators. This behaviour arises due to the weaker coupling strength between the resonators, as reflected in the reduced values of the coupling coefficient $\Omega$ with respect to the effective loss of the system, shown in Fig. \ref{f:5}(e).
\begin{figure}[!ht]
	\centering
	\includegraphics[width=0.6\linewidth]{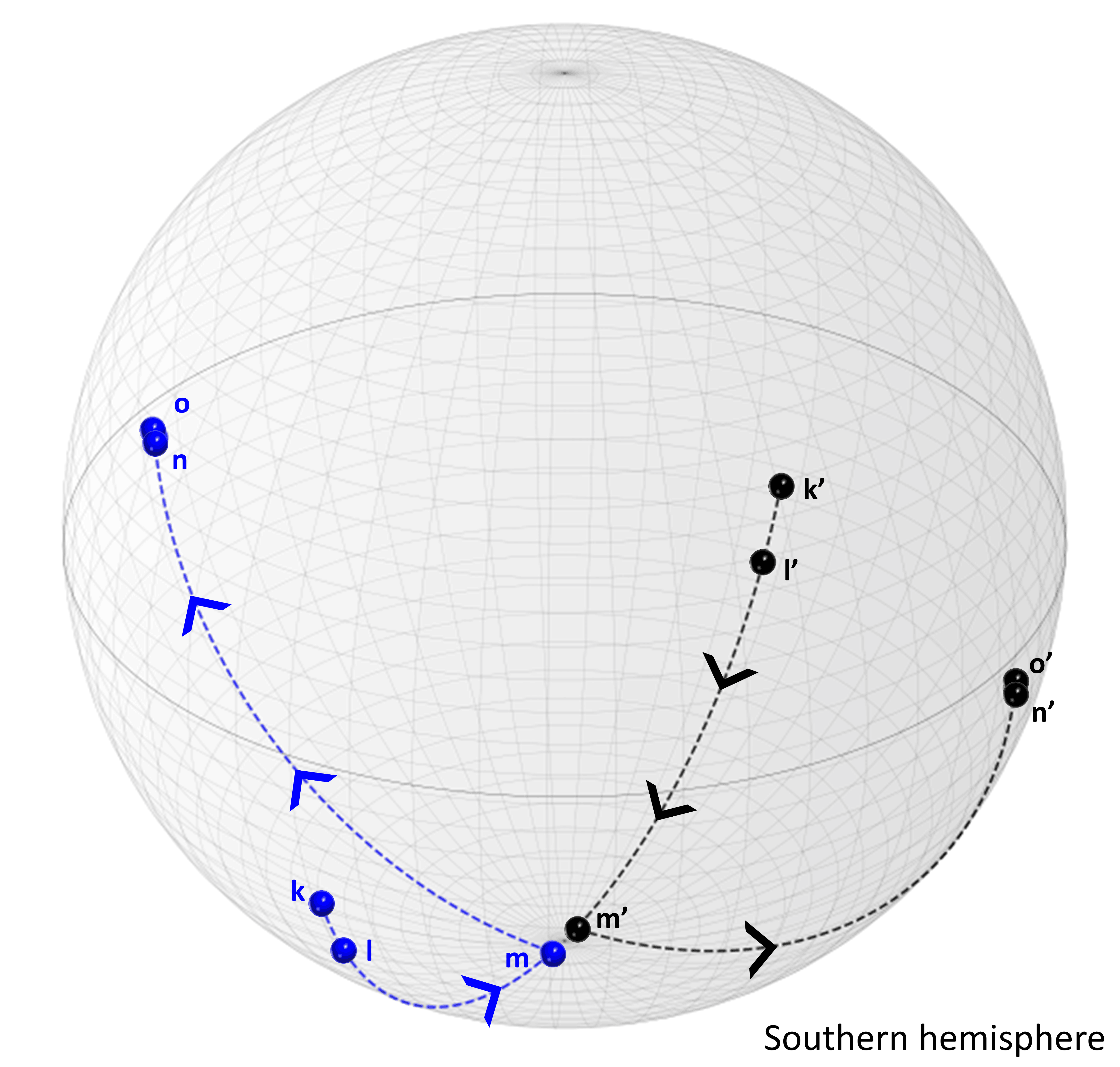}
	\caption{\footnotesize Polarization eigenstates for different S values plotted on Poincaré sphere which exhibit all eigenstates are located in southern hemisphere, the dashed lines represent the eigenstates of an ideal PT symmetric plasmonic metasurface with varying coupling coefficient. The arrows show the transition from (k $\&$ k’) to (o $\&$ o’), depicting the eigenstates of PT symmetry at the southern hemisphere for S${_\text{1,2,3,4,5}}$ = 64 to 72 µm, respectively, where (m $\&$ m’) represent eigenstates for exceptional point at S${_\text{3}}$= 68 µm near south pole. }
	\label{f:7}       
\end{figure}
\section{Poincaré sphere for PT symmetric eigenpolarization states}
To address the non-Hermitian nature of the metasurface, the eigen-polarization states for various inter-cavity separation distances S are plotted on Poincaré sphere’s surface, as shown in Fig. \ref{f:7}. All elliptical eigen-polarization states lie on the surface of the southern hemisphere, depicting the transition from PT-symmetric states (k, l and k', l') to PT-asymmetric states (n, o and n', o') via the exceptional point at S$_3$=68 µm (m and m'). At this point, the states m and m' nearly coincide at the south pole, indicating the near coalescence of the eigenstates.
The confinement of all eigenpolarization states to a single hemisphere confirms PT-symmetric behaviour at the exceptional point and highlights the non-Hermitian nature of the plasmonic meta-structure.

\section{Conclusion}
Parity-time (PT) symmetry, including its breaking through the exceptional point (EP), has been experimentally examined in a non-Hermitian THz plasmonic metasurface comprising orthogonally oriented circular complementary split-ring resonators. The terahertz time domain spectroscopy technique has been used to investigate the transmission properties of the proposed novel PT symmetric meta-design. The PT-symmetric nature of the metasurface is investigated by systematically varying the inter-cavity separation distances S${_\text{1,2,3,4,5}}$ = 64, 66, 68, 70, and 72 µm, effectively displacing one cavity relative to the second cavity. Eigenvalue spectra, eigenphase evolution, and eigenpolarization states were measured at each separation to track the PT symmetric phase transition using the coupled mode theory applied to our geometry. The system preserves a PT-symmetric phase up to S${_\text{3}}$=68 µm, where an exceptional point is observed at around 0.567 THz. Beyond this critical point, a sharp transition into the PT-broken (asymmetric) phase occurs, driven by a reduction in near-field coupling between the resonators. To further explore the evolution of the PT symmetric system’s eigenstates and non-Hermitian nature of the meta-geometry, Poincaré sphere representations are employed. The observation again reveals that the metasurface undergoes a transition from the PT-symmetric phase to the PT-asymmetric phase from the exceptional point. This study presents a novel demonstration of PT symmetry and exceptional point dynamics in a THz plasmonic metasurface. The findings hold significant promise for developing next-generation THz photonic components, including ultra-sensitive sensors, tunable filters, perfect absorbers, polarization and phase modulators, holographic devices, etc.

\section{Disclosures} 
The authors declare no conflicts of interest.
\section{Data Availability Statement}
The data supporting the findings of this study are available from the corresponding author upon reasonable request.

\section{Acknowledgments}
The authors MI and GK would like to acknowledge financial support from the Anusandhan National Research Foundation (formerly Science and Engineering Research Board) of the Government of India for research grant No. (CRG/2023/000392). The authors AP and GK gratefully acknowledge financial support from the SERB, India, (CRG/2021/002187).\\
Authors acknowledge Shaivyaa Mehndiratta for helping with drafting the manuscript. And archive the manuscript.

\section*{References\label{bibby}}
\bibliographystyle{elsarticle-num}
\bibliography{article1}

\end{document}